RESEARCH

# Discovery of a Physically Interpretable Data-Driven Wind-Turbine Wake Model

**Kherlen Jigjid[1] · Ali Eidi[1] · Nguyen Anh Khoa Doan[1] · Richard P. Dwight[1]**

## Abstract
This study presents a compact data-driven Reynolds-averaged Navier-Stokes (RANS) model for wind turbine wake prediction, built as an enhancement of the standard $k$-$\varepsilon$ formulation. Several candidate models were discovered using the symbolic regression framework Sparse Regression of Turbulent Stress Anisotropy (SpaRTA), trained on a single Large Eddy Simulation (LES) dataset of a standalone wind turbine. The leading model was selected by prioritizing simplicity while maintaining reasonable accuracy, resulting in a novel linear eddy viscosity model. This selected leading model reduces eddy viscosity in high-shear regions—particularly in the wake—to limit turbulence mixing and delay wake recovery. This addresses a common shortcoming of the standard $k$-$\varepsilon$ model, which tends to overpredict mixing, leading to unrealistically fast wake recovery. Moreover, the formulation of the leading model closely resembles that of the established $k$-$\varepsilon$-$f_P$ model. Consistent with this resemblance, the leading and $k$-$\varepsilon$-$f_P$ models show nearly identical performance in predicting velocity fields and power output, but they differ in their predictions of turbulent kinetic energy. In addition, the generalization capability of the leading model was assessed using three unseen six-turbine configurations with varying spacing and alignment. Despite being trained solely on a standalone turbine case, the model produced results comparable to LES data. These findings demonstrate that data-driven methods can yield interpretable, physically consistent RANS models that are competitive with traditional modeling approaches while maintaining simplicity and achieving generalizability.

✉ Kherlen Jigjid
k.jigjid@tudelft.nl

✉ Richard P. Dwight
r.p.dwight@tudelft.nl

[1] Aerodynamics, Faculty of Aerospace Engineering, TU Delft, Kluyverweg 2, HT Delft 2629, Netherlands

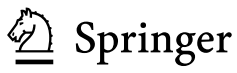

## 1 Introduction

Wake prediction is essential for optimizing wind farm layout, aiming to maximize energy output while minimizing land use and turbine loading caused by turbulence. In large wind farms, wakes can reduce the total energy yield by 10–20% (Barthelmie et al. 2009). Furthermore, turbulence generated by wakes can induce structural vibrations in downstream turbines, leading to premature fatigue and increased maintenance costs (Réthoré 2009).

Computational Fluid Dynamics (CFD) methods, such as Large Eddy Simulation (LES) and Reynolds-averaged Navier-Stokes (RANS) simulation, are commonly employed for wake studies. LES resolves large-scale turbulence while modeling smaller scales, providing results that align well with measurement data (Troldborg et al. 2011). However, its high computational cost of about $10^3$–$10^4$ CPU hours (Porté-Agel et al. 2019), driven by fine temporal and spatial resolution requirements, limits its practicality for widespread applications. In contrast, RANS solves steady-state flow fields while modeling all turbulence scales, enabling the use of coarser meshes and significantly reducing computational costs. The penalty is that accuracy in separated and wake flows is generally poor, with e.g. wake recovery being mispredicted (Réthoré 2009).

Among RANS models, the standard $k$-$\varepsilon$ model (Launder and Spalding 1974) is widely used in turbine wake prediction. However, in wake regions, it tends to significantly overpredict turbulent viscosity, leading to excessive wake mixing and too rapid recovery (Réthoré 2009; Sanderse et al. 2011; van der Laan 2014; Eidi et al. 2021). This motivated various modifications to the model. For instance, El Kasmi and Masson (2008) introduced an additional term in the near-turbine region to suppress Turbulent Kinetic Energy (TKE) production, improving wake prediction with subsequent fine-tuning of model constants (Réthoré 2009). Similarly, Zehtabiyan-Rezaie and Abkar (2024) derived an analytical term accounting for turbine-induced forces in the TKE transport equation, indirectly reducing eddy viscosity and delaying wake recovery.

The above modifications fall within the framework of Linear Eddy Viscosity Models (LEVMs) and thus cannot fully capture the anisotropy of the Reynolds Stress Tensor (RST). Indeed, the standard $k$-$\varepsilon$ model is known to violate RST realizability conditions in wake regions (Réthoré 2009). This limitation stems from the Boussinesq eddy viscosity assumption, which linearly relates RST to the mean strain rate tensor through eddy viscosity.

To overcome this, Nonlinear Eddy Viscosity Models (NLEVMs) extend the relationship to include non-linear terms. Van der Laan (2014) investigated different NLEVMs (Taulbee 1992; Apsley and Leschziner 1998) for single turbine wake prediction. While these models could achieve improved velocity and RST anisotropy predictions, they faced stability issues due to their higher-order terms. A key finding was that the velocity prediction improvement primarily came from an auxiliary term designed to limit unbounded growth of the coefficients in the model, rather than from the non-linear terms themselves. This term reduces eddy viscosity in high shear regions like wakes. This finding led to the development of the $k$-$\varepsilon$-$f_P$ model (van der Laan et al. 2015), which is obtained by simplifying a cubic NLEVM (Apsley and Leschziner 1998) by retaining only the linear term. However, it was also seen that the model overestimates TKE in the far wake and does not capture the RST anisotropy correctly. Nevertheless, for velocity profile prediction of wakes, anisotropy capture is not necessary. The reason is velocity recovery is primarily driven by the gradient of the off-diagonal components of the RST (van der Laan et al. 2023), which can be predicted by the $k$

-$\varepsilon$-$f_P$ model as reported in Baungaard et al. (2022). Considering this, the model has proven effective for accurately predicting velocity deficits and power output for multiple turbines in a row, despite over-predicting TKE (Eidi et al. 2021).

Beyond traditional model modifications, data-driven approaches have recently emerged as an alternative. One such method is the Sparse Regression of Turbulent Stress Anisotropy (SpaRTA) framework by Schmelzer et al. (2019). It introduces two extra algebraic terms into two-equation models, such as the $k$-$\varepsilon$ model. These terms are based on Pope's non-linear tensor basis and their invariants (Pope 1975), and correct both the RST anisotropy and TKE transport equation. A key benefit of this approach compared to neural-network based data-driven models, is that it yields models for the corrective terms from high-fidelity data (e.g., time-averaged LES or measurement data) consisting of explicit algebraic expressions, making them more easily interpretable.

SpaRTA has been used for wake modeling in prior studies (Steiner et al. 2020, 2022, 2022). The discovered models were trained on setups with a small number of turbines and generalize well to scenarios involving multiple wake interactions, including cases with yawed turbines. However, due to a prioritization of accuracy, the resulting models are relatively complex. They consist of dozens of terms with activation switches based on local flow conditions, leading to numerical instability and limited interpretability. Consequently, there remains a need to explore the framework's potential for discovering models that prioritize simplicity—thereby aiding interpretability.

In this study, we address this challenge by utilizing the SpaRTA framework to discover RANS wake models that prioritize interpretability and stability through the use of the fewest possible terms. As a baseline, we adopt the standard $k$-$\varepsilon$ model. Models for the RST correction are discovered from LES data of a single-turbine scenario, and the one with the fewest terms is selected. Notably, the selected data-driven model exhibits a formulation and behavior very similar to the established $k$-$\varepsilon$-$f_P$ model in that they both work as eddy viscosity limiters, despite being derived independently. Consequently, the $k$-$\varepsilon$-$f_P$ model is used as a benchmark to evaluate the performance of the data-driven model's predictions of streamwise velocity and TKE fields through analysis of its eddy viscosity fields. Furthermore, to test the generalization capability of the data-driven model beyond its training data, its performance is assessed on three unseen six-turbine wind farm layouts with various levels of wake interactions. This assessment includes streamwise velocity, TKE, and power predictions.

The remainder of this paper is structured as follows: Sect. 2 reviews the $k$-$\varepsilon$-$f_P$ model and compares it with the standard $k$-$\varepsilon$ model, followed by the implementation details of the SpaRTA framework. This section concludes by describing the wind farm layouts and training data used in our study. Section 3 presents the discovered model and analyzes its similarity to the $k$-$\varepsilon$-$f_P$ model. The model's performance is then evaluated across various turbine arrangements to validate its generalization capabilities. Finally, Sect. 4 concludes the paper with a summary of findings and implications for future research.

## 2 Methodology

### 2.1 RANS modelling, standard $k$-$\varepsilon$, and $k$-$\varepsilon$-$f_P$ models

In this study, the $k$-$\varepsilon$-$f_P$ model is selected as the benchmark for evaluating the data-driven model (introduced later in Sect. 3.1), as both models share a similar formulation that incorporates an eddy viscosity limiter into the standard $k$-$\varepsilon$ model. Therefore, understanding the benchmark model is essential for interpreting the behavior of the data-driven model.

This section presents the $k$-$\varepsilon$-$f_P$ model (van der Laan 2014) through its formulation and highlights the limitations of the standard $k$-$\varepsilon$ model by comparing simulation results for a single-turbine wake. First, the governing equations for RANS simulations are introduced, together with the standard $k$-$\varepsilon$ model equations. Subsequently, the $k$-$\varepsilon$-$f_P$ model is presented as a modification of the standard $k$-$\varepsilon$ model. Finally, the formulations are examined, and the two models are compared in terms of their predictions for the streamwise velocity and TKE fields.

#### 2.1.1 The $k$-$\varepsilon$ Model

To perform the simulations, we solve the incompressible RANS equations, augmented by an Atmospheric Boundary Layer (ABL) driving force $f^{\mathrm{ABL}}$ and an Actuator Disk (AD) model force $f^{\mathrm{AD}}$:

$$\frac{\partial U_i}{\partial x_i} = 0, \tag{1}$$

$$U_j \frac{\partial U_i}{\partial x_j} = -\frac{1}{\rho}\frac{\partial p}{\partial x_i} + \frac{\partial}{\partial x_j}\left(2\nu S_{ij} - \tau_{ij}\right) + \delta_{ix}\left(f^{\mathrm{ABL}} + f^{\mathrm{AD}}\right). \tag{2}$$

Here, $U_i$ represents the mean velocity in the $i$-direction, $\rho$ is the density, $p$ is the mean pressure, $\nu$ is the molecular kinematic viscosity, $S_{ij} = \frac{1}{2}\left(\frac{\partial U_i}{\partial x_j} + \frac{\partial U_j}{\partial x_i}\right)$ is the mean strain rate tensor, and $\tau_{ij}$ denotes the RST. To conduct RANS simulations using the above governing equations, closure modeling is required to approximate $\tau_{ij}$.

In the standard $k$-$\varepsilon$ model, $\tau_{ij}$ is approximated using the Boussinesq eddy viscosity assumption:

$$\tau_{ij} := -2\nu_t S_{ij} + \frac{2}{3}k\delta_{ij}, \tag{3}$$

where $k$ is the TKE, and $\nu_t$ is the eddy viscosity defined as:

$$\nu_t := C_\mu \frac{k^2}{\varepsilon}. \tag{4}$$

The coefficient $C_\mu$ is a model constant and $\varepsilon$ is the TKE dissipation rate. For high shear flows, the standard $k$-$\varepsilon$ model tends to overpredict $k$, which leads to amplified $\nu_t$. This results in excessive turbulent mixing, causing faster wake recovery than is observed in reality (Réthoré 2009; Sanderse et al. 2011; van der Laan 2014; Eidi et al. 2021).

#### 2.1.2 The $k$-$\varepsilon$-$f_P$ Model

To limit the overestimated $\nu_t$, the $k$-$\varepsilon$-$f_P$ model modifies the standard $k$-$\varepsilon$ model by incorporating a local flow-dependent eddy viscosity limiter $f_P$, while keeping other model equations unchanged. The modified eddy viscosity is expressed as:

$$\nu_t^* := C_\mu f_P \frac{k^2}{\varepsilon}. \tag{5}$$

This formulation can be interpreted as employing an effective $C_\mu^* = C_\mu f_P$ that adapts to local flow conditions. The term $f_P$ is defined using a local shear parameter $\sigma = \frac{k}{\varepsilon}\sqrt{\left(\frac{\partial U_i}{\partial x_j}\right)^2}$ as follows:

$$f_P\left(\sigma/\widetilde{\sigma}\right) := \frac{2f_0}{1+\sqrt{1+4f_0\left(f_0-1\right)\left(\sigma/\widetilde{\sigma}\right)^2}}, \quad f_0 := \frac{C_R}{C_R-1}. \tag{6}$$

The ratio $\sigma/\tilde{\sigma}$ quantifies the deviation of local flow from the log-law regime by comparing it with the reference shear parameter $\widetilde{\sigma} = C_\mu{}^{-0.5}$, which is obtained from calibration under log-law conditions. The model uses $C_\mu = 0.03$, which is typical for atmospheric flows (Richards and Hoxey 1993). The wake recovery is controlled by the Rotta constant $C_R = 4.5$, tuned using eight LES wind turbine datasets in the original study (van der Laan 2014).

The relation (6) exhibits different behaviors depending on the local flow conditions. In wake regions where shear is high $\sigma > \tilde{\sigma}$, $f_P < 1$. This reduction in $f_P$ decreases $\nu_t^*$, which delays wake recovery. Far from wake-affected regions or in the inflow where $\sigma = \tilde{\sigma}$, $f_P = 1$, and the model behaves like the standard $k$-$\varepsilon$ model with $C_\mu = 0.03$. Note that $f_P > 1$ can occur when $\sigma < \tilde{\sigma}$, although such conditions are typically limited to slip Boundary Condition (BC) regions—where the velocity gradient approaches zero—and to localized regions within turbine wakes. The function $f_P$ is bounded, reaching its maximum value of $9/7$ at $\sigma = 0$ and approaching 0 as $\sigma \to \infty$; thus, $0 < f_P < 9/7$.

#### 2.1.3 Prediction Comparison Between the $k$-$\varepsilon$ and $k$-$\varepsilon$-$f_P$ Models

The performance of the $k$-$\varepsilon$-$f_P$ and standard $k$-$\varepsilon$ models is evaluated by comparing their predictions with LES data for a single turbine wake in Fig. 1. For this comparison, the $C_\mu$ value in the $k$-$\varepsilon$ model was set to 0.03 to align with that of the $k$-$\varepsilon$-$f_P$ model. Both RANS simulations were conducted with the parameters in Table 1 (for the RANS simulation configurations, refer to App. A in the supplementary material). Profiles in the figure are scaled by the inlet velocity at the turbine hub height, $U_h = 8$ m/s. The $k$-$\varepsilon$-$f_P$ model shows improved streamwise velocity $U_x$ predictions with delayed wake recovery. This stems from the $f_P$ limiter, which reduces the excessive $\nu_t$ and consequently decreases turbulence mixing. Regarding the $k$ fields, both models predict unphysically high values, but the $k$-$\varepsilon$-$f_P$ model better matches with the LES, especially in the near-wake region. Such overprediction of $k$ may undermine the $k$-$\varepsilon$-$f_P$ model's reliability for blade load calculations, though it remains well-suited for power prediction applications requiring accurate $U_x$ profiles.

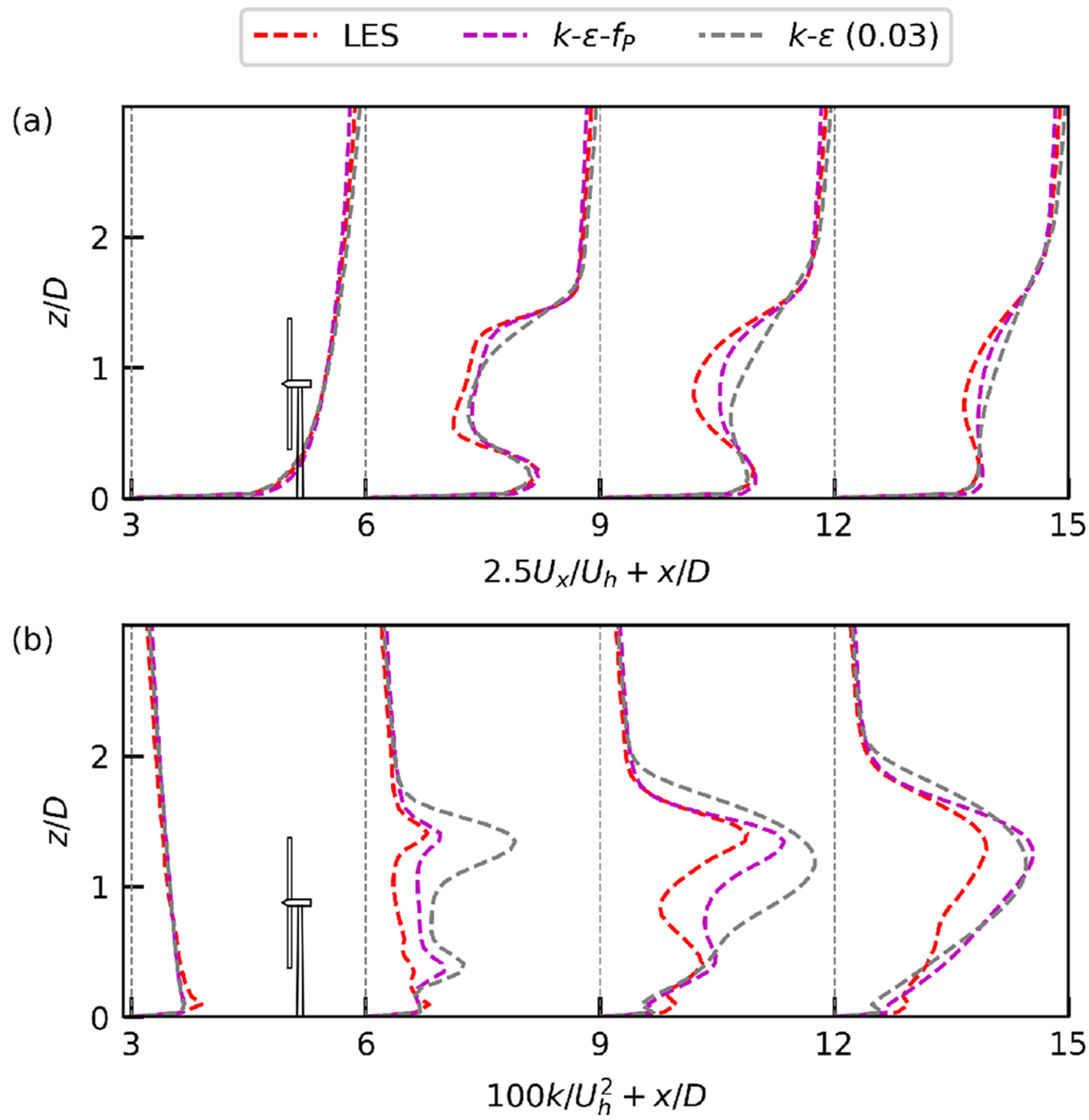

**Fig. 1** Comparison of LES, $k-\varepsilon-f_P$ and $k-\varepsilon$ models for prediction of (**a**) $U_x$ and (**b**) $k$ profiles. The plots show the $x$-$z$ plane at $y = 2.5D$ for the single turbine case 1T in Fig. 2

**Table 1** Turbulence model parameters

| Turbulence model | $C_\mu$ | $C_{\varepsilon 1}$ | $C_{\varepsilon 2}$ | $\sigma_\varepsilon$ | $\sigma_k$ | $z_0$ [m] | $f^{\mathrm{ABL}}$ [m/s$^2$] |
|---|---|---|---|---|---|---|---|
| Propagation & data-driven | 0.09 | 1.42 | 1.92 | 1.00 | 1.30 | 2.98e-3 | 2.90e-4 |
| $k$-$\varepsilon$-$f_P$ | 0.03 | 1.21 | 1.92 | 1.30 | 1.00 | 0.67e-3 | 2.03e-4 |
| $k$-$\varepsilon$ ($C_\mu = 0.03$) | 0.03 | 1.42 | 1.92 | 1.00 | 1.30 | 1.50e-3 | 2.01e-4 |
| $k$-$\varepsilon$ ($C_\mu = 0.09$) | 0.09 | 1.42 | 1.92 | 1.00 | 1.30 | 11.39e-3 | 3.48e-4 |

## 2.2 Data-Driven RANS Modeling for Wakes: SpaRTA

The SpaRTA framework (Schmelzer et al. 2019) is applied in this study to obtain a RANS model for wake flows. Specifically, it is used to derive an algebraic correction term for the RST in the standard $k$-$\varepsilon$ model. The standard $k$-$\varepsilon$ model was chosen as the baseline model for the SpaRTA framework due to its simplicity, widespread use in the field, and its role as the foundation for various extended models (including the $k$-$\varepsilon$-$f_P$ model).

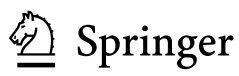

The following subsections detail the implementation of the framework, covering the incorporation of corrective terms into the baseline model, their extraction from reference LES data, the model discovery process for the corrective terms, and simulation methods used to assess their effects.

### 2.2.1 Introducing Corrective Terms for the Baseline Model

The SpaRTA framework starts with introducing two corrective terms to the baseline model, which is the standard $k$-$\varepsilon$ model in our case. The *first* corrective term, $b_{ij}^{\Delta}$, modifies turbulence anisotropy in the deviatoric part of RST as:

$$\tau_{ij}^{c} := 2k\left(b_{ij} + b_{ij}^{\Delta} + \frac{1}{3}\delta_{ij}\right), \quad \text{where} \quad b_{ij} := -\frac{\nu_t}{k}S_{ij}, \tag{7}$$

compared with (3).

The *second* corrective term, $R$, acts as a residual term in the $k$-equation. It has a similar effect as the production term and is therefore also included in the $\varepsilon$-equation for consistency. As a result, the $k$- and $\varepsilon$-equations become:

$$\frac{Dk}{Dt} = (P_k^c + R) - \varepsilon + \frac{\partial}{\partial x_j}\left[\left(\nu + \frac{\nu_t}{\sigma_k}\right)\frac{\partial k}{\partial x_j}\right], \tag{8}$$

$$\frac{D\varepsilon}{Dt} = C_{\varepsilon 1}\left(P_k^c + R\right)\frac{\varepsilon}{k} - C_{\varepsilon 2}\frac{\varepsilon^2}{k} + \frac{\partial}{\partial x_j}\left[\left(\nu + \frac{\nu_t}{\sigma_\varepsilon}\right)\frac{\partial \varepsilon}{\partial x_j}\right]. \tag{9}$$

Here, the coefficients $C_\mu$, $C_{\varepsilon 1}$, $C_{\varepsilon 2}$, $\sigma_\varepsilon$, and $\sigma_k$ are model constants, and the specific values used in this work are listed in Table 1. The term $P_k^c$ represents the modified production rate of TKE due to the term $b_{ij}^{\Delta}$, and is defined as:

$$P_k^c := -\tau_{ij}^{c}\frac{\partial U_i}{\partial x_j} = -\tau_{ij}\frac{\partial U_i}{\partial x_j} + P_k^{\Delta}, \quad \text{where} \quad P_k^{\Delta} := -2k\, b_{ij}^{\Delta}\frac{\partial U_i}{\partial x_j}. \tag{10}$$

For wake prediction applications, the corrective terms introduced above are further decomposed into ABL- and wake-related components. In order to match the RANS model of the ABL to the LES reference, corrective terms are also introduced into the undisturbed ABL. These corrections, denoted $b_{ij}^{\Delta,\mathrm{ABL}}$ and $R^{\mathrm{ABL}}$, take the same form as the wake corrections (i.e. modifying the turbulence anisotropy, and the $k$-budget), except that they are treated as a function of wall-distance only, see Steiner et al. (2022) and Jigjid et al. (2024) for details. The total corrections applied are then:

$$b_{ij}^{\Delta} := b_{ij}^{\Delta,\mathrm{ABL}}(\mathbf{z}) + b_{ij}^{\Delta,\mathrm{W}}, \tag{11}$$

$$R := R^{\mathrm{ABL}}(\mathbf{z}) + R^{\mathrm{W}}. \tag{12}$$

This decomposition enables model discovery for wake-related corrections ($b_{ij}^{\Delta,\mathrm{W}}$ and $R^{\mathrm{W}}$) only. Meanwhile, the optimal ABL terms are directly used—with no model—and are inlet condition dependent.

### 2.2.2 Extracting Optimal Corrective Fields from LES Data

After incorporating the corrective terms $b_{ij}^{\Delta,\mathrm{ABL}}$, $b_{ij}^{\Delta,\mathrm{W}}$, $R^{\mathrm{ABL}}$ and $R^{\mathrm{W}}$ into the baseline model equations, the next step is to determine their optimal values. In the SpaRTA framework, this is achieved using the *frozen approach*, in which high-fidelity LES statistics—$\tilde{U}_i$, $\tilde{k}$, and $\tilde{\tau}_{ij}$—are injected into the RANS Eqs. (1)-(4) and (7)-(10). In this approach, the LES-derived fields are treated as fixed inputs—hence the term "frozen"—which allows the equations to be solved for the unknown corrective terms and thus obtain their optimal values. The tilde notation used in this study, $\tilde{\cdot}$, denotes fields obtained from LES data and, for clarity, does not represent the filtering operator commonly used in the LES context.

To clarify how these LES statistics are derived: $\tilde{U}_i$ is computed by time-averaging the instantaneous LES velocity field $u_i$ over a defined interval, i.e., $\tilde{U}_i = \langle u_i \rangle$, where $\langle \cdot \rangle$ denotes time-averaging. The fluctuating velocity is then $u_i' = u_i - \tilde{U}_i$, from which the TKE is calculated as $\tilde{k} = \langle u_i' u_i' \rangle / 2$, and the RST as $\tilde{\tau}_{ij} = \langle u_i' u_j' \rangle$. In this way, the LES statistics are prepared for use in the SpaRTA framework.

For wind farm applications, both ABL- and wake-related corrections in (11) and (12) are obtained in a sequential manner. First, the ABL-related corrections $\tilde{b}_{ij}^{\Delta,\mathrm{ABL}}(\mathbf{z})$ and $\tilde{R}^{\mathrm{ABL}}(\mathbf{z})$ are obtained using a one-dimensional inlet profile simulations (for details, see App. B in the supplementary material). After determining these ABL corrections, we keep them fixed in (11) and (12) and apply the frozen approach to calculate the wake-related corrective terms $\tilde{b}_{ij}^{\Delta,\mathrm{W}}(\mathbf{x})$ and $\tilde{R}^{\mathrm{W}}(\mathbf{x})$, defined as functions of the spatial coordinate $\mathbf{x} \in \mathbb{R}^3$. As such the LES solution satisfies the RANS equations, given the discovered corrections.

After obtaining the optimal values for the corrective terms, a verification step, called *propagation*, validates the corrective terms by implementing them in the RANS solver. The corrective terms are deemed verified when the resulting velocity and TKE fields closely match the LES data, making them suitable targets for the subsequent model discovery phase.

### 2.2.3 Model Discovery for the Corrective Terms

Having determined the optimal values of the corrective terms, we then proceed to identify their functional expressions using a sparse symbolic regression approach.

To achieve a perfect match between LES reference and RANS predictions, the corrective terms $\tilde{b}_{ij}^{\Delta}$ for turbulence anisotropy and $\tilde{R}$ for the kinetic energy budget are required, including both ABL- and wake-related components. However, in this study we observe empirically that the majority of the improvement in the velocity field stems from $\tilde{b}_{ij}^{\Delta,\mathrm{W}}$, while $\tilde{R}^{\mathrm{W}}$ offers only minor improvements to the TKE in the wake. This is consistent with the observations of Réthoré (2009), that turbulence anisotropy is the leading source of error in turbine wakes for Boussinesq models. Additionally, focusing on $\tilde{b}_{ij}^{\Delta,\mathrm{W}}$ enables the development of a simpler modification of the baseline model, aligning with the main objective of the research. Regarding the ABL-related corrective terms, we do not attempt to model them, as they are inlet-specific and thus case-dependent. Instead, we directly use the optimal values.

Following Pope's general effective-viscosity formulation (Pope 1975), we assume that $b_{ij}$ can be represented solely in terms of local velocity gradients. Consequently, the tensor $b_{ij}^{\Delta,\mathrm{W}}$ can be expressed as:

$$b_{ij}^{\Delta,\mathrm{W}} := \sum_{n=1}^{10} \alpha_n(I_1, I_2, I_3, I_4, I_5) T_{ij}^{(n)}, \tag{13}$$

where $T_{ij}^{(n)}$ are the base tensors and $I_m$ are the corresponding invariants (for definitions of $T_{ij}^{(n)}$ and $I_m$, see Pope (1975)). Since both $T_{ij}^{(n)}$ and $I_m$ are constructed from components of $\frac{\partial U_i}{\partial x_j}$, the model discovery procedure reduces to identifying the functional coefficients $\alpha_n$.

To address the unknown structure of $\alpha_n$, we assume that $b_{ij}^{\Delta,\mathrm{W}}$ can be expressed as a linear combination of different terms, that are constructed using $T_{ij}^{(n)}$, $I_m$, and a set of functions (specifically, $(\cdot)^{-1}$, $(\cdot)^{0.5}$, $(\cdot)^{1}$, $(\cdot)^{2}$, and $\tanh(\cdot)$ in this study). Each term is formed by combining a single $I_m$ with one of the functions and then multiplying the result by one of $T_{ij}^{(n)}$. This process generates a total of 220 candidate terms, and the corrective term is expressed using these terms:

$$\hat{b}_{ij}^{\Delta,\mathrm{W}} := C^T \Theta = \left[ T_{ij}^{(1)},\ I_1 T_{ij}^{(1)},\ I_1^{0.5} T_{ij}^{(1)},\ \ldots,\ \tanh(I_5) T_{ij}^{(10)} \right] \begin{bmatrix} \theta_1 \\ \theta_2 \\ \theta_3 \\ \vdots \\ \theta_{220} \end{bmatrix}. \tag{14}$$

Here, $\Theta$ denotes the vector of coefficients corresponding to each candidate term in the function library $C$. In this formulation, the model is discovered by determining the optimal set of coefficients in $\Theta$. Please note that, to prioritize discovering simple models, we used only unary operators when constructing the function library (14), meaning each term involves at most one invariant. Including binary operators—which would allow terms to combine two invariants—would increase the library from 220 to 2320 terms and introduce additional complexity, so this was deemed unnecessary for the present study.

We pose the problem of finding an optimal $\Theta$ as a least-squares minimization augmented with regularization terms to promote sparsity. Specifically, we incorporate Elastic Net (EN) penalties (Zou and Hastie 2005), formulated as:

$$\Theta^{\mathrm{EN}} := \underset{\Theta}{argmin} \left( \| \tilde{b}_{ij}^{\Delta,\mathrm{W}} - C^T \Theta \|_2^2 + \underbrace{\alpha\rho \| \Theta \|_1}_{L_1 \text{ norm}} + \underbrace{\frac{\alpha(1-\rho)}{2} \| \Theta \|_2^2}_{L_2 \text{ norm}} \right), \tag{15}$$

where the penalties combine the $L_1$ and $L_2$ norms of the coefficient vector. The $L_1$ norm encourages sparsity by driving smaller coefficients toward zero, whereas the $L_2$ norm shrinks larger coefficients, limiting their overall magnitudes. The intensity of regularization is governed by the parameter $\alpha$, while the blending between the two penalties is controlled by the mixing parameter $\rho$. A range of models is explored by varying the values of $\alpha$ and $\rho$. For each resulting $\Theta^{\mathrm{EN}}$, coefficients below predefined thresholds are discarded, along with their associated functions in $C$, to further promote sparsity. The remaining coefficients are then refined using a least-squares formulation similar to (15), but without the $L_1$ norm pen-alty, thereby resembling Ridge regression (Hoerl and Kennard 1970). This re-optimization yields the final sparse model, characterized by minimal number of terms.

### 2.2.4 Evaluation of Corrective Terms in RANS Simulations

To investigate the effectiveness of the corrective terms, this study considers three different RANS simulations under the following conditions:

- *Full propagation RANS*: The frozen corrections, $\tilde{b}_{ij}^{\Delta,\mathrm{W}}$ and $\tilde{R}^{\mathrm{W}}$, obtained by injecting the LES data into the baseline model, are used directly. This represents the best possible RANS prediction, but is only computable given LES data for the same case.
- *RST propagation RANS*: The frozen corrections for $\tilde{b}_{ij}^{\Delta,\mathrm{W}}$ are used, but not those for $\tilde{R}^{\mathrm{W}}$. Given that, in the following, we will search for models only for $\tilde{b}_{ij}^{\Delta,\mathrm{W}}$, this represents the best possible outcome of our model predictions.
- *Data-driven RANS*: A trained model is used to predict $\hat{b}_{ij}^{\Delta,\mathrm{W}}$, demonstrating the performance of the data-driven model obtained.

Besides the wake-related corrective terms, all simulations incorporate frozen corrective terms for the ABL, $\tilde{b}_{ij}^{\Delta,\mathrm{ABL}}$ and $\tilde{R}^{\mathrm{ABL}}$. The corrective terms used in each simulation are listed in Table 2. Further details on the RANS configurations are provided in App. A, and the parameters used for each model are tabulated in Table 1.

## 2.3 Wind Farm Layouts and Reference LES Dataset

This section describes the wind farm layouts employed for both model development and the assessment of its generalizability. Additionally, the LES datasets used as reference cases are introduced.

### 2.3.1 Wind Farm Layouts

For this study, various wind farm layouts are considered, as shown in Fig. 2. These cases are designed to represent varying levels of wake complexity:

- Case 1T: A single turbine, representing the simplest wake scenario with no wake interactions.
- Case 6T5D: Six turbines aligned with $5D$ spacing, creating the most intense wake interactions due to the close proximity of the turbines.
- Case 6T7D: Six turbines aligned with $7D$ spacing, resulting in reduced wake interactions compared to the $5D$ configuration.

**Table 2** Corrective terms used for RANS simulations

| Turbulence model | $R =$ | $b_{ij}^{\Delta} =$ |
|---|---|---|
| Full propagation | $\tilde{R}^{\mathrm{ABL}} + \tilde{R}^{\mathrm{W}}$ | $\tilde{b}_{ij}^{\Delta,\mathrm{ABL}} + \tilde{b}_{ij}^{\Delta,\mathrm{W}}$ |
| RST propagation | $\tilde{R}^{\mathrm{ABL}}$ | $\tilde{b}_{ij}^{\Delta,\mathrm{ABL}} + \tilde{b}_{ij}^{\Delta,\mathrm{W}}$ |
| Data-driven | $\tilde{R}^{\mathrm{ABL}}$ | $\tilde{b}_{ij}^{\Delta,\mathrm{ABL}} + \hat{b}_{ij}^{\Delta,\mathrm{W}}$ |

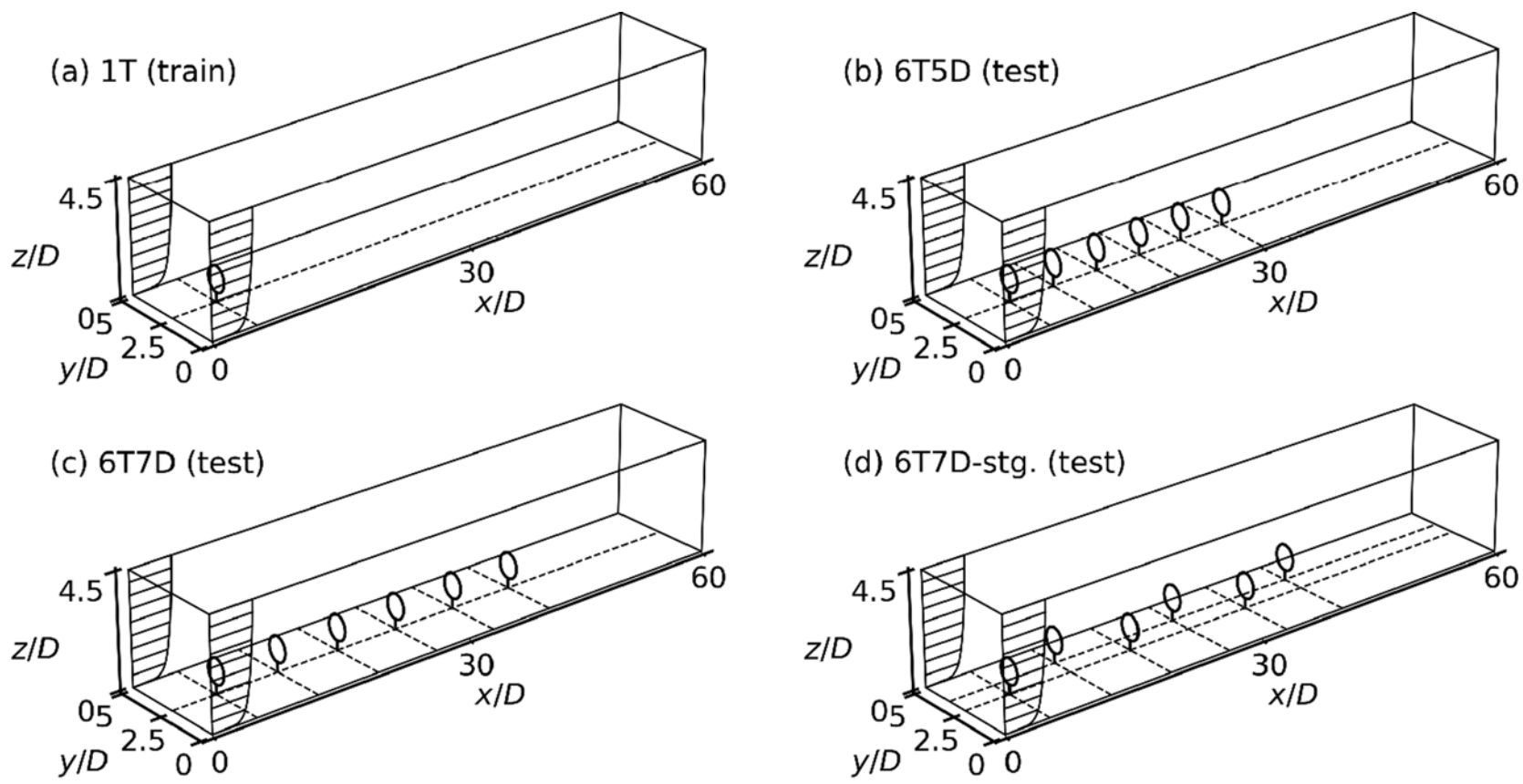

**Fig. 2** Wind farm layouts for the cases: (**a**) 1T, (**b**) 6T5D, (**c**) 6T7D, and (**d**) 6T7D-staggered. The lengths along each axis are normalized by the rotor diameter $D$

- Case 6T7D-staggered: Six turbines with $7D$ spacing, where even-numbered turbines are shifted by $1D$ in the $y$-direction, introducing partial wake interactions and a more realistic wind farm layout.

Only case 1T is used for training the model, representing the simplest wake scenario with no wake interactions. To evaluate the generalizability of the discovered model, the three additional wind farm layouts are used as test cases. These layouts are selected based on the availability of LES datasets from Eidi et al. (2021, 2022).

### 2.3.2 LES Dataset

The reference LES datasets used in this study are succeeded from earlier works (Eidi et al. 2021, 2022), which were generated using a pseudo-spectral in-house code described in Porté-Agel et al. (2011) and Abkar and Porté-Agel (2015). The simulations employ a Lagrangian scale-dependent dynamic Smagorinsky Subgrid Scale (SGS) model (Porté-Agel et al. 2000; Wu and Porté-Agel 2015), a widely validated approach for wind farm simulations, particularly through comparisons with wind tunnel experiments (Wu and Porté-Agel 2011, 2013).

For turbine parameterization in the LES simulations, a non-rotating AD model was employed. This model performs well in large-scale wind farm simulations for capturing the dominant flow features (Stevens et al. 2018), and it can also be implemented in the RANS framework due to its suitability for steady-state simulations. This makes the AD model a suitable choice for the simulations in our study. While more detailed models, such as the Actuator Line (AL) method, can resolve near-wake structures (e.g., tip and hub vortices), they require significantly finer grids and unsteady simulations, making them unsuitable for this study.

### 2.4 Selection of Training Points

To reduce the computational cost of model discovery and improve the representativeness of the training set, training points are sub-sampled from the wake region rather than using all available samples. The full propagation RANS simulation of the single turbine case (1T) serves as the training data, and only points where the corrective term $\tilde{b}_{ij}^{\Delta,\mathrm{W}}$ is active are selected. Specifically, these are defined by the condition $|\tilde{P}_k^{\Delta,\mathrm{W}}| > 0.001$, where

$$\tilde{P}_k^{\Delta,\mathrm{W}} := -2k\,\tilde{b}_{ij}^{\Delta,\mathrm{W}}\frac{\partial U_i}{\partial x_j}. \tag{16}$$

In Fig. 3, the spatial distribution of $\tilde{P}_k^{\Delta,\mathrm{W}}$ is shown. Noticeable negative values appear at the edges of the turbine wake region, indicating TKE mitigation due to the corrective term. This threshold-based selection primarily retains samples from the wake and near-wall regions, as highlighted by the contour lines. However, near-wall samples are excluded from the training set due to their sensitivity to wall function modeling. In addition, the threshold value of 0.001 was chosen based on visual inspection and may require adjustment for different simulation setups.

Using this wake-based sampling, SpaRTA successfully converged for the RST correction model. In contrast, sampling from the entire domain led to convergence issues during model discovery. However, neither of the two sampling approaches—wake-only or full-domain—resulted in a converged model for the $R$ term.

## 3 Findings

### 3.1 Selection of the Leading Model from the Discovered Data-Driven Models

In this section, we present the data-driven models obtained in the current study and outline the selection process of a leading model. These models are derived using the SpaRTA framework (Sect. 2.2) with different combinations of $\alpha$ and $\rho$ in (15), applied to the training samples (Sect. 2.4) extracted from the propagation RANS simulation of case 1T. The leading model is selected based on its simplicity, prioritizing a lower number of terms to enhance interpretability. The selected model is then used for comparison with the bench-mark $k$-$\varepsilon$-$f_P$ model to evaluate the effectiveness of the SpaRTA framework for wake model discovery in the following sections.

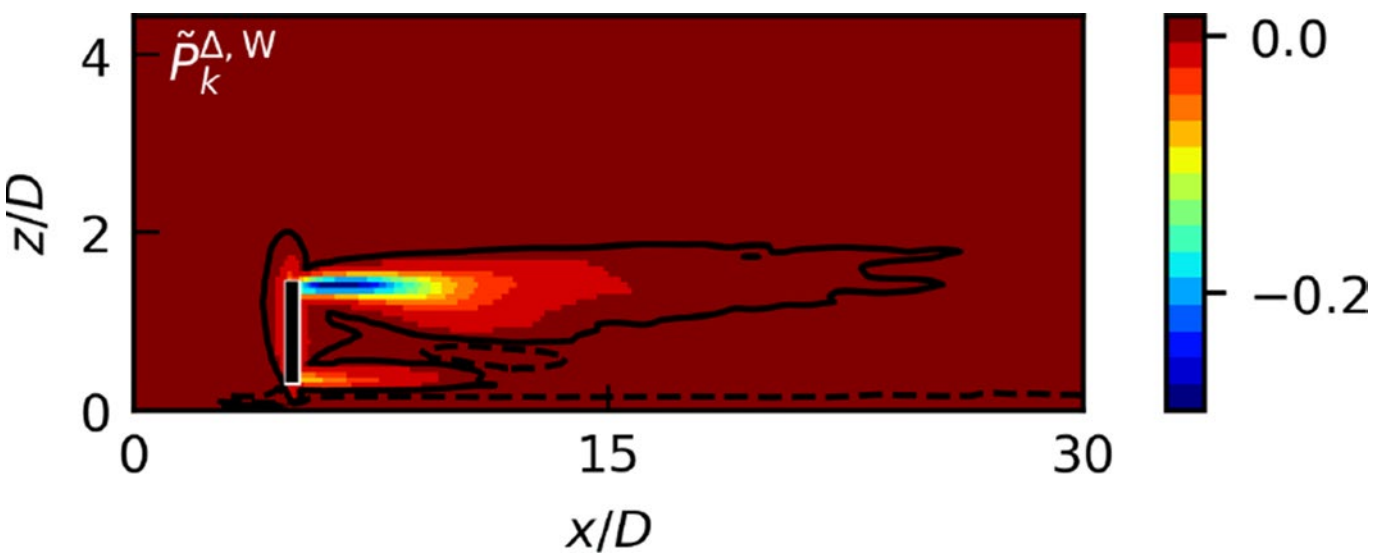

**Fig. 3** The $\tilde{P}_k^{\Delta,\mathrm{W}}$ field obtained from the full propagation RANS for the case 1T (training data). The plot shows the $x$-$z$ plane at $y = 2.5D$. The solid and dashed lines represent −0.001 and 0.001, respectively

#### 3.1.1 Discovered Data-Driven Models

The obtained data-driven models achieve coefficients of determination $R^2$ between 0.82 and 0.89 on the training samples, with varying features and number of terms. Table 3 presents the five simplest models, ordered by term count. More complex models, including the most complex one (13 terms, $R^2 = 0.89$), are omitted to maintain focus on simpler formulations, despite their higher $R^2$.

As shown in the table, the term "$\tanh(I_1)T_{ij}^{(1)}$" appears in all models and has the largest contribution to $R^2$, highlighting the importance of including both $I_1$ and $T_{ij}^{(1)}$ in the model. In contrast, additional terms provide only marginal improvements in $R^2$. Based on this, we focus on the first two models—the single- and two-term models—which are the simplest and consist solely of $I_1$ and $T_{ij}^{(1)}$, to select the leading data-driven model.

#### 3.1.2 Selection of the Leading Data-Driven Model

To select the leading model from the two potential models mentioned above, we compare their eddy viscosity limiter formulations. The $b_{ij}^{\Delta,\mathrm{W}}$ formulations for the single- and two-term models are, respectively:

$$\hat{b}_{ij}^{\Delta,\mathrm{W,I}} := \frac{1}{11.99}\tanh\left(\frac{I_1}{81.10}\right)T_{ij}^{(1)}, \tag{17}$$

$$\hat{b}_{ij}^{\Delta,\mathrm{W,II}} := \frac{1}{68.01}T_{ij}^{(1)} + \frac{1}{14.96}\tanh\left(\frac{I_1}{81.10}\right)T_{ij}^{(1)}. \tag{18}$$

Here, $\hat{\cdot}$ denotes quantities predicted by the data-driven model, and the Roman numeral superscript indicates whether the model is a single- (I) or two-term (II). Both models share a common tensor basis $T_{ij}^{(1)} = S_{ij}\frac{k}{\varepsilon}$ in every term, allowing $S_{ij}$ to be factored out when (17) or (18) replace $b_{ij}^{\Delta}$ in (7). This allows us to reformulate them similarly to the eddy viscosity limiter in the $k$-$\varepsilon$-$f_P$ model as:

**Table 3** List of discovered data-driven models

| Model name | Term number | $R^2$ | Included terms |
|---|---|---|---|
| Single-term | 1 | 0.82 | $\tanh(I_1)T_{ij}^{(1)}$ |
| Two-term | 2 | 0.82 | $\tanh(I_1)T_{ij}^{(1)}$, $T_{ij}^{(1)}$ |
| Three-term | 3 | 0.84 | $\tanh(I_1)T_{ij}^{(1)}$, $T_{ij}^{(1)}$, $I_5^{-1}T_{ij}^{(2)}$ |
| Four-term | 4 | 0.83 | $\tanh(I_1)T_{ij}^{(1)}$, $T_{ij}^{(1)}$, $I_1^{0.5}T_{ij}^{(1)}$, $\tanh(I_2)T_{ij}^{(1)}$ |
| Five-term | 5 | 0.84 | $\tanh(I_1)T_{ij}^{(1)}$, $T_{ij}^{(1)}$, $I_1^{0.5}T_{ij}^{(1)}$, $\tanh(I_2)T_{ij}^{(1)}$, $I_5^{-1}T_{ij}^{(2)}$ |

$$\hat{f}_P^{\mathrm{I}} = 1 - \frac{1}{C_\mu}\frac{1}{11.99}\tanh\left(\frac{I_1}{81.10}\right), \tag{19}$$

$$\hat{f}_P^{\mathrm{II}} = 1 - \frac{1}{C_\mu}\frac{1}{68.01} - \frac{1}{C_\mu}\frac{1}{14.96}\tanh\left(\frac{I_1}{81.10}\right), \tag{20}$$

where $I_1 = T_{ij}^{(1)}T_{ji}^{(1)}$, which is always positive, represents the local shear. As a result, the predicted effective eddy viscosity coefficient $\hat{C}_\mu^* = C_\mu \hat{f}_P$ is reduced in high-shear regions, reflecting the same limiting behavior as the original $f_P$.

Complementary to the formulation comparison, we compare the $\hat{f}_P^{\mathrm{I}}$ and $\hat{f}_P^{\mathrm{II}}$ fields computed from the full propagation RANS simulation data of the 1T case (training data), as shown in Fig 4. Both models show nearly identical $\hat{f}_P$ in low-value regions corresponding to the wake, as reflected in the similar shapes of the $\hat{f}_P = 0.5$ contour lines (black). However, outside the wake region, the two fields differ: $\hat{f}_P^{\mathrm{II}}$ shows lower values compared to $\hat{f}_P^{\mathrm{I}}$. For example, the region enclosed by the $\hat{f}_P = 0.7$ contour (grey) expands in the $\hat{f}_P^{\mathrm{II}}$ field, indicating that undervaluation begins to occur in higher $\hat{f}_P$ regions. Moreover, the $\hat{f}_P^{\mathrm{II}}$ field lacks the $\hat{f}_P = 0.9$ contour (white), suggesting a substantial reduction of $C_\mu$ across the entire domain.

This difference arises from the additional constant term in the $\hat{f}_P^{\mathrm{II}}$ formulation, which reduces its values regardless of local flow conditions. In contrast, the single-term model depends solely on the local invariant $I_1$, making it fully local. Therefore, since both models achieve the same $R^2$, we selected the single-term model for further comparison with the $k$-$\varepsilon$-$f_P$ model.

### 3.2 Eddy Viscosity Limiter Comparison of the Data-Driven and the $k$-$\varepsilon$-$f_P$ Models

In this section, we compare the selected data-driven model, specifically the single-term model in the previous section, to the $k$-$\varepsilon$-$f_P$ model to investigate their similarities. First, we examine their mathematical formulations, followed by a comparison of the eddy viscosity limiter fields $f_P$ and $\hat{f}_P^{\mathrm{I}}$ obtained from the corresponding RANS simulations for the single-turbine layout 1T.

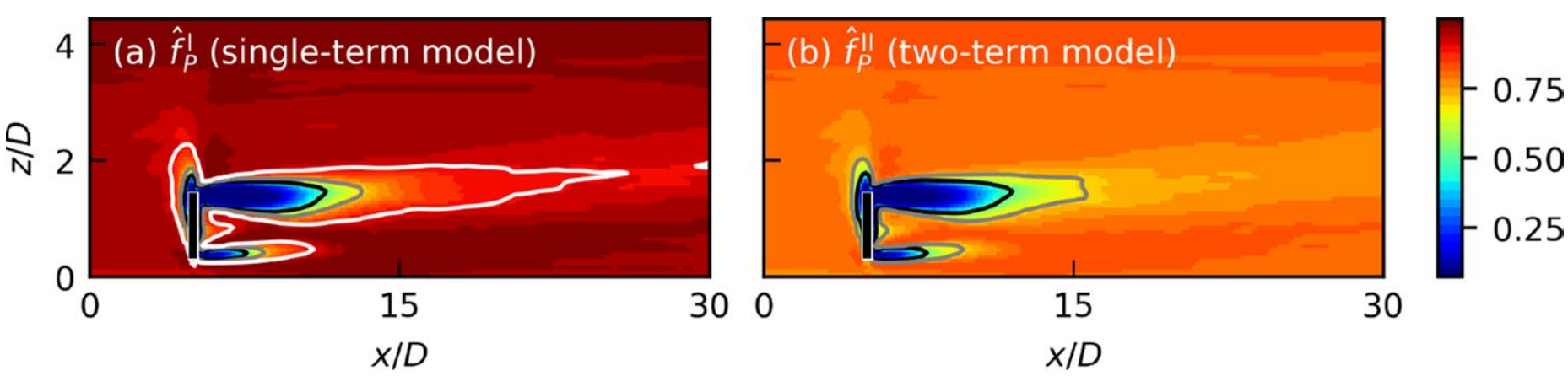

**Fig. 4** Calculated $\hat{f}_P$ fields using the full propagation RANS data (case 1T): (**a**) single-term model, (**b**) two-term model. View at $y = 2.5D$ with contour lines at 0.5 (black), 0.7 (grey), and 0.9 (white)

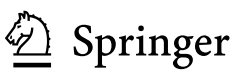

### 3.2.1 Formulation Comparison

Having the explicit symbolic formulations allows us to interpret and understand model behavior—an important advantage of the symbolic regression approach over black-box methods. In the following, we leverage this to analyze and compare the formulations of both models.

The formulations differ in two key aspects. First, the boundary values of the limiters differ. Although it is physically unrealistic, $\hat{f}_P^{\mathrm{I}}$ approaches a minimum value of approximately 0.073 as $I_1 \to \infty$, when the "$\tanh$" converges to unity. This calculation uses $C_\mu = 0.09$, consistent with the value used in the simulation. Conversely, in regions with no shear $I_1 = 0$, $\hat{f}_P^{\mathrm{I}}$ reaches its maximum value of 1.0. Thus, the single-term model is bounded by $0.073 < \hat{f}_P^{\mathrm{I}} < 1.0$, which differs from the bounds of the $k$-$\varepsilon$-$f_P$ model with $C_\mu = 0.03$, $0 < f_P < 9/7$, as mentioned earlier in Sect. 2.1. Nevertheless, the negative correlation between the limiter and local shear is a common feature shared by both models.

Second, the models differ in their choice of input variables as seen in their formulas (6) and (19). The single-term model utilizes $I_1$, while the $k$-$\varepsilon$-$f_P$ model employs $\sigma$ to define the local shear level. The relationship between these variables is:

$$\sigma = \left(I_1 + \omega_{ij}\omega_{ij}\right)^{1/2}, \tag{21}$$

where $\omega_{ij} = \frac{1}{2}\left(\frac{\partial U_i}{\partial x_j} - \frac{\partial U_j}{\partial x_i}\right)\frac{k}{\varepsilon}$ is the mean rotation rate tensor that is non-dimensionalized by turbulent scale, and for $\omega_{ij}\omega_{ij} << I_1$ regions, the relation becomes $\sigma \approx (I_1)^{1/2}$.

To compare the relationship between the two variables, their distributions are calculated using the training data fields and shown in Fig. 5. In the wake region, the magnitudes differ, likely due to the presence of rotation, yet their spatial distributions remain nearly identical across the domain. The Pearson correlation coefficient between the two fields was found to be 0.99. These comparisons suggest that although the two variables are not exactly the same, they are very similar, implying that the models rely on nearly identical flow quantities.

### 3.2.2 Spatial Distribution Comparison

In addition to the formula-based analysis, we directly compare the $\hat{f}_P^{\mathrm{I}}$ and $f_P$ fields for the 1T layout using results from RANS simulations with the respective models, as shown in Fig. 6, to verify whether the conclusions drawn above still hold.

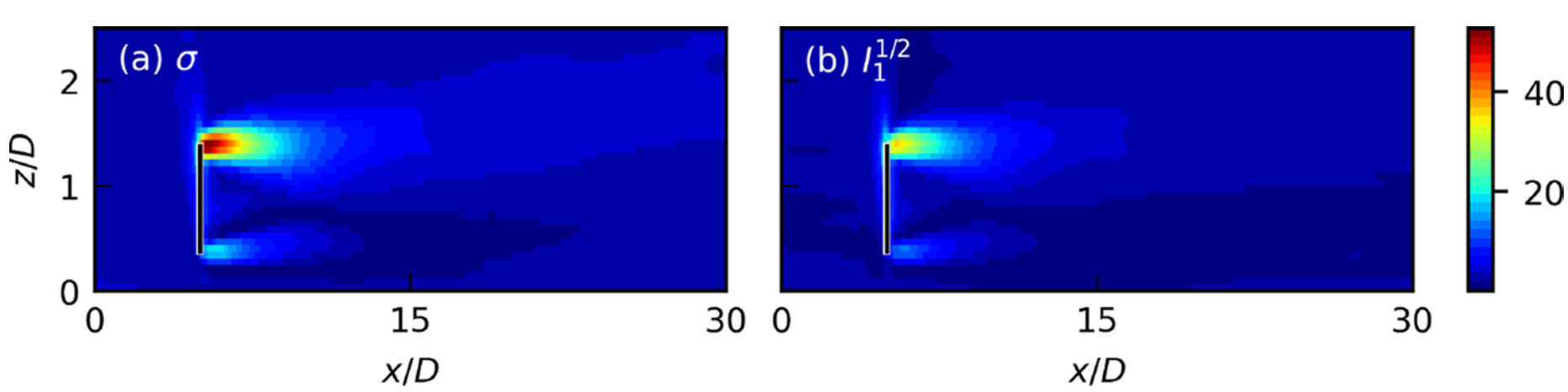

**Fig. 5** Calculated (**a**) $\sigma$ and (**b**) $I_1^{1/2}$ fields using the full propagation RANS data (case 1T). View at $y = 2.5D$

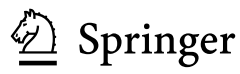

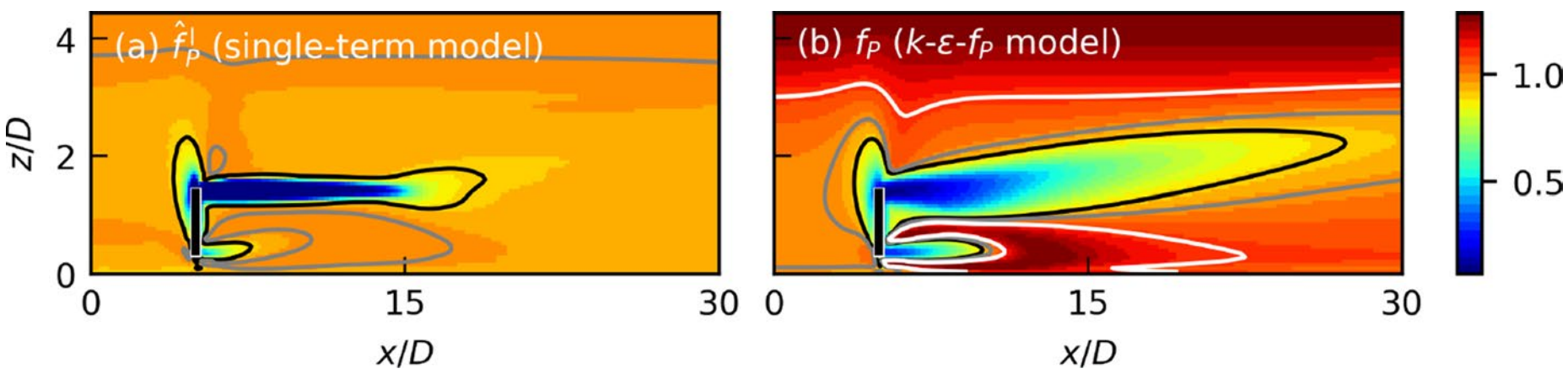

**Fig. 6** Predicted (**a**) $\hat{f}^{\mathrm{I}}_P$ and (**b**) $f_P$ fields from the corresponding RANS simulations (case 1T). View at $y = 2.5D$ with contour lines at 0.9 (black), 0.98 (grey), and 1.1 (white)

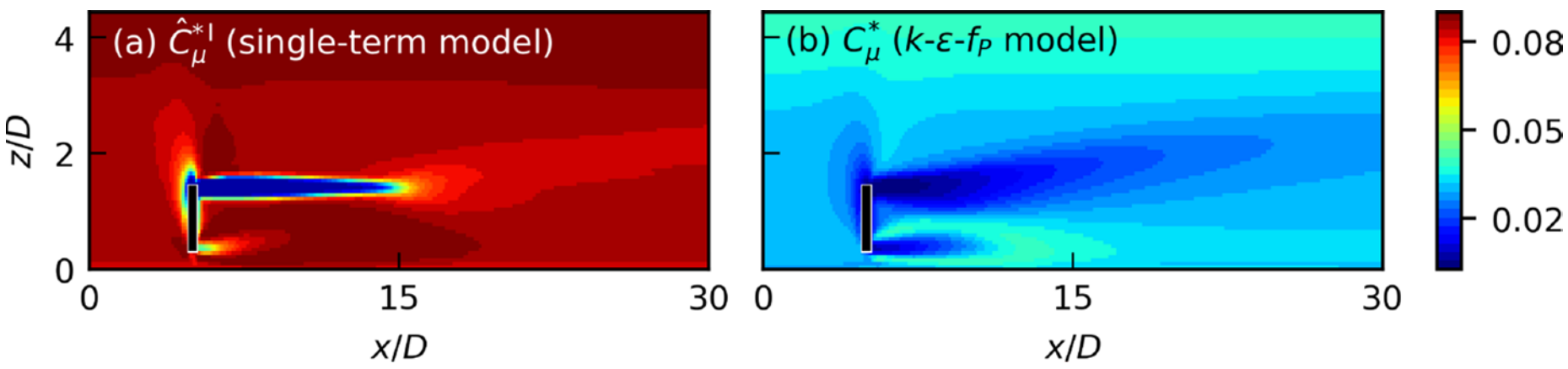

**Fig. 7** Predicted (**a**) $\hat{C}^{*\,\mathrm{I}}_\mu$ and (**b**) $C^*_\mu$ fields from the corresponding RANS simulations (case 1T). View at $y = 2.5D$

Both fields exhibit similarly low values in the wake region, delineated by the black contour line representing 0.9. The single-term model produces a compact region with sharp gradients, whereas the $k$-$\varepsilon$-$f_P$ model yields a more smoothly distributed field. Despite these differences in distribution, both models substantially limit the eddy viscosity in the region, contributing to the prediction of delayed wake recovery.

In other parts of the domain, the $k$-$\varepsilon$-$f_P$ model predicts $f_P > 1.0$ in certain areas. This overestimation is particularly evident near the top boundary (slip condition) and in a small enclosed region behind the turbine, indicated by the white contour line ($f_P > 1.1$). These regions correspond to the condition $\sigma < \tilde{\sigma}$. In contrast, the single-term model does not produce regions where $\hat{f}^{\mathrm{I}}_P > 1.0$. Notably, the regions where $\hat{f}^{\mathrm{I}}_P > 0.98$ closely align with areas where $f_P > 1.1$, suggesting that while both models identify similar high-value regions, the single-term model caps values under 1.0.

The above comparisons demonstrate that both models predict remarkably similar eddy viscosity limiter fields, consistent with the formulation analysis. However, their actual influence on simulations occurs through $\hat{C}^{*\,\mathrm{I}}_\mu$ and $C^*_\mu$. Since these quantities derive from different $C_\mu$ values, they show substantially different magnitudes while preserving the spatial distribution patterns of the $\hat{f}^{\mathrm{I}}_P$ and $f_P$ fields, as plotted in Fig. 7. Consequently, despite the similarity in eddy viscosity limiter fields, the models can be expected to predict distinctive flow fields.

### 3.2.3 Summary

In this section, we compared the selected data-driven model to the $k$-$\varepsilon$-$f_P$ model, focusing on their eddy viscosity limiter formulations. Both models exhibit similar behavior, charac-terized by a reduction of eddy viscosity in high-shear regions, and rely on closely related flow quantities as input variables. However, their limiting behavior differs due to the distinct capping values in their formulations. These similarities and differences were consistently reflected in the eddy viscosity limiter fields predicted by the corresponding RANS simulations. Nevertheless, despite the close agreement in the limiter fields, the actual impact on the simulations is expected to differ significantly due to the different $C_\mu$ values employed in each model.

### 3.3 Effect of Eddy Viscosity Modifications on Wake Prediction

In the previous section, we demonstrated that the single-term model acts as an eddy viscosity limiter. This section examines how the modified $\nu_t$ affects flow prediction performance.

First, we evaluate the effectiveness of optimal corrective terms derived from LES data to establish the expected improvements from these corrections. Subsequently, we examine the single-term model's predicted $\nu_t$ field to assess its impact on the $U_x$ and $k$ predictions. This is conducted through comparison with several reference cases: results from the RANS models described in Sect. 2.2.4 and the $k$-$\varepsilon$-$f_P$ model, and the reference LES data.

#### 3.3.1 Effectiveness of Corrective Terms

The effectiveness of LES-derived optimal corrective terms is assessed by comparing RANS simulations incorporating these terms with the original LES data and baseline models. Figure 8 compares prediction results from various RANS models with reference LES data for the 1T wind farm layout. The left, middle, and right columns represent the $U_x$, $k$, and $\nu_t$ fields, respectively. The first row displays the LES data, rows two through four show the RANS models outlined in Sect. 2.2.4, and the final three rows present existing models, including $k$-$\varepsilon$ models with varying $C_\mu$ values, as indicated in parentheses.

Among the models shown, the full propagation RANS simulation, incorporating the corrective terms $\tilde{b}_{ij}^{\Delta,\mathrm{W}}$ and $\tilde{R}^{\mathrm{W}}$, achieves $U_x$ and $k$ fields that closely align with the LES results. This represents a significant improvement over the baseline $k$-$\varepsilon$ model with $C_\mu = 0.09$, highlighting the effectiveness of the corrective terms derived from the LES data using the frozen approach.

Notably, the RST propagation RANS simulation, using only the corrective term $\tilde{b}_{ij}^{\Delta,\mathrm{W}}$, yields a $U_x$ field nearly identical to both the full propagation simulation and the LES data, with slightly faster wake recovery. The $k$ and $\nu_t$ fields show minor differences but remain closely aligned with the full propagation results. These confirms that $\tilde{b}_{ij}^{\Delta,\mathrm{W}}$ is the primary contributor to the improved prediction, supporting the focus on modeling only this term, which also simplifies the model.

#### 3.3.2 Assessing Model Performance Through Eddy Viscosity

Having established the effectiveness of the corrective terms, we now turn to evaluating the single-term model's performance, with primary comparison to the $k$-$\varepsilon$-$f_P$ model. This evaluation focuses on how both models predict eddy viscosity, and how it affects the flow fields.

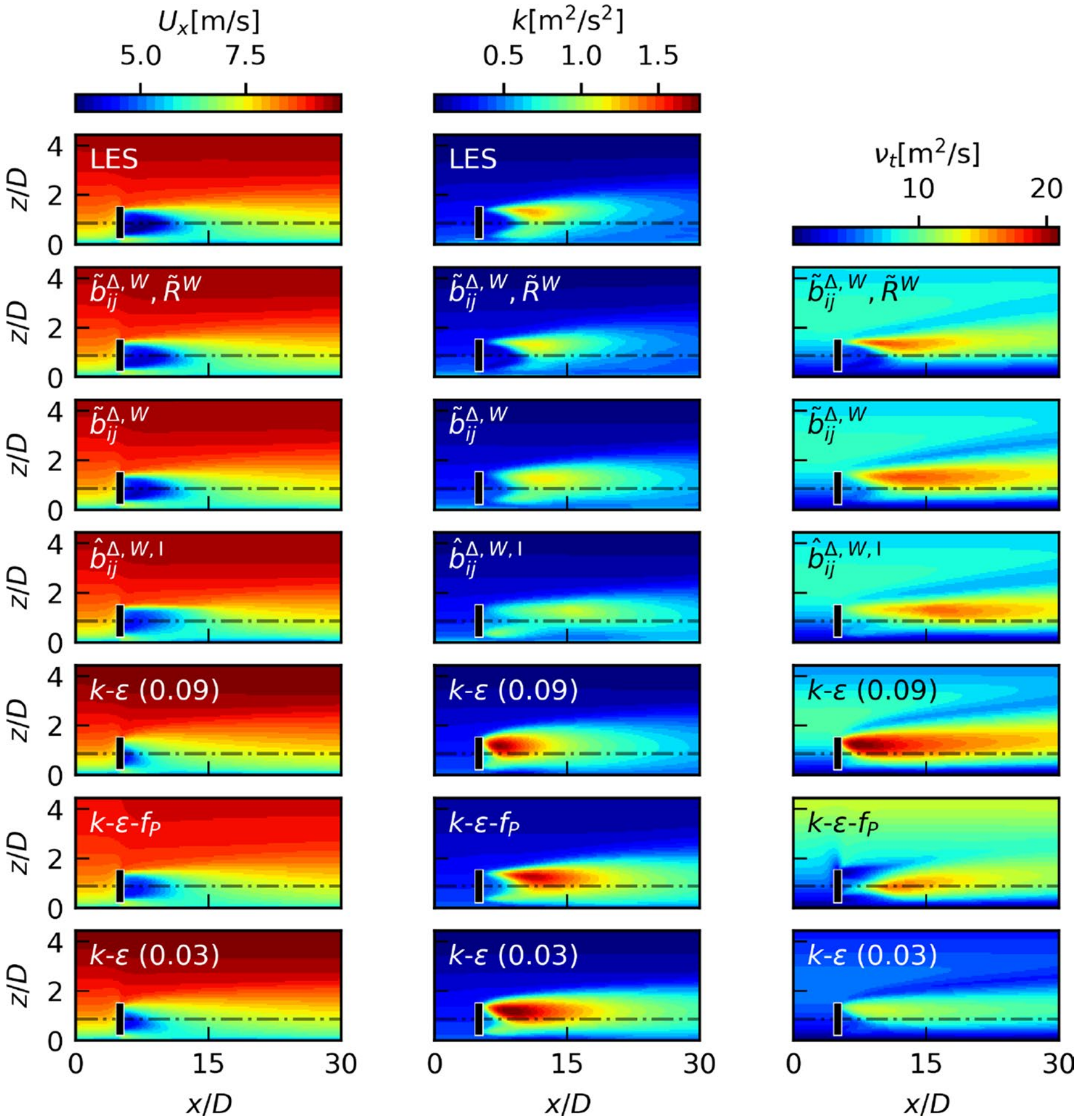

**Fig. 8** Comparison of LES and various RANS models for prediction of $U_x$, $k$ and $\nu_t$ fields for the case 1T. The plots show the $x$-$z$ plane at $y = 2.5D$

In Fig. 8, both the single-term and $k$-$\varepsilon$-$f_P$ models generate $U_x$ fields that closely align with the LES data, demonstrating significant improvement in wake prediction compared to their respective baseline $k$-$\varepsilon$ models. However, their predictions for $k$ fields differ substantially. The $k$-$\varepsilon$-$f_P$ model considerably overpredicts TKE in wake regions relative to LES data, while the data-driven model underpredicts it. When compared to their baselines, the $k$-$\varepsilon$-$f_P$ model maintains a similar $k$ distribution pattern, though with the high-$k$ region slightly shifted downstream and reduced in magnitude. In contrast, the single-term model exhibits markedly lower $k$ predictions than its baseline. This presents an interesting question: how do these models produce nearly identical $U_x$ distributions despite such different $k$ predictions?

This can be explained by examining the $\nu_t$ predictions. For the baseline models, where $C_\mu$ is constant, the $\nu_t$ fields exhibit distributions highly similar to those of $k$. However, the single-term and $k$-$\varepsilon$-$f_P$ models lack this spatial correlation due to the effects of $\hat{C}_\mu^{*\,\mathrm{I}}$ and $C_\mu^*$, respectively. Both models significantly reduce $\nu_t$ immediately behind the turbine at its top edge, thereby decreasing turbulence mixing compared to their respective baselines. This region is vital for the wake recovery process, as it is where mixing between the free stream and the wake occurs. In the far wake region, both models allow slower mixing with the freestream, leading to gradual wake recovery. In this manner, both models effectively adjust $\nu_t$ independently of $k$, limiting turbulence mixing and yielding wake recovery patterns that more closely match the LES data.

Evidence of this $\nu_t$ reduction in the near wake region is clearly visible in the profiles shown in Fig. 9. At $x = x_\mathrm{T} + 1D$, the single-term and $k$-$\varepsilon$-$f_P$ models predict significantly lower $\nu_t$ values near the upper wake boundary compared to their respective baseline models (comparing black to cyan lines, and grey to magenta lines). Furthermore, both models exhibit more varied $\nu_t$ profiles, which contrasts with the shape of the baseline models' pro-files that closely reflect the $k$ distribution. Additionally, the $U_x$ field reveals that the baseline models exhibit recovery already underway due to excessive mixing, even in the near-wake region.

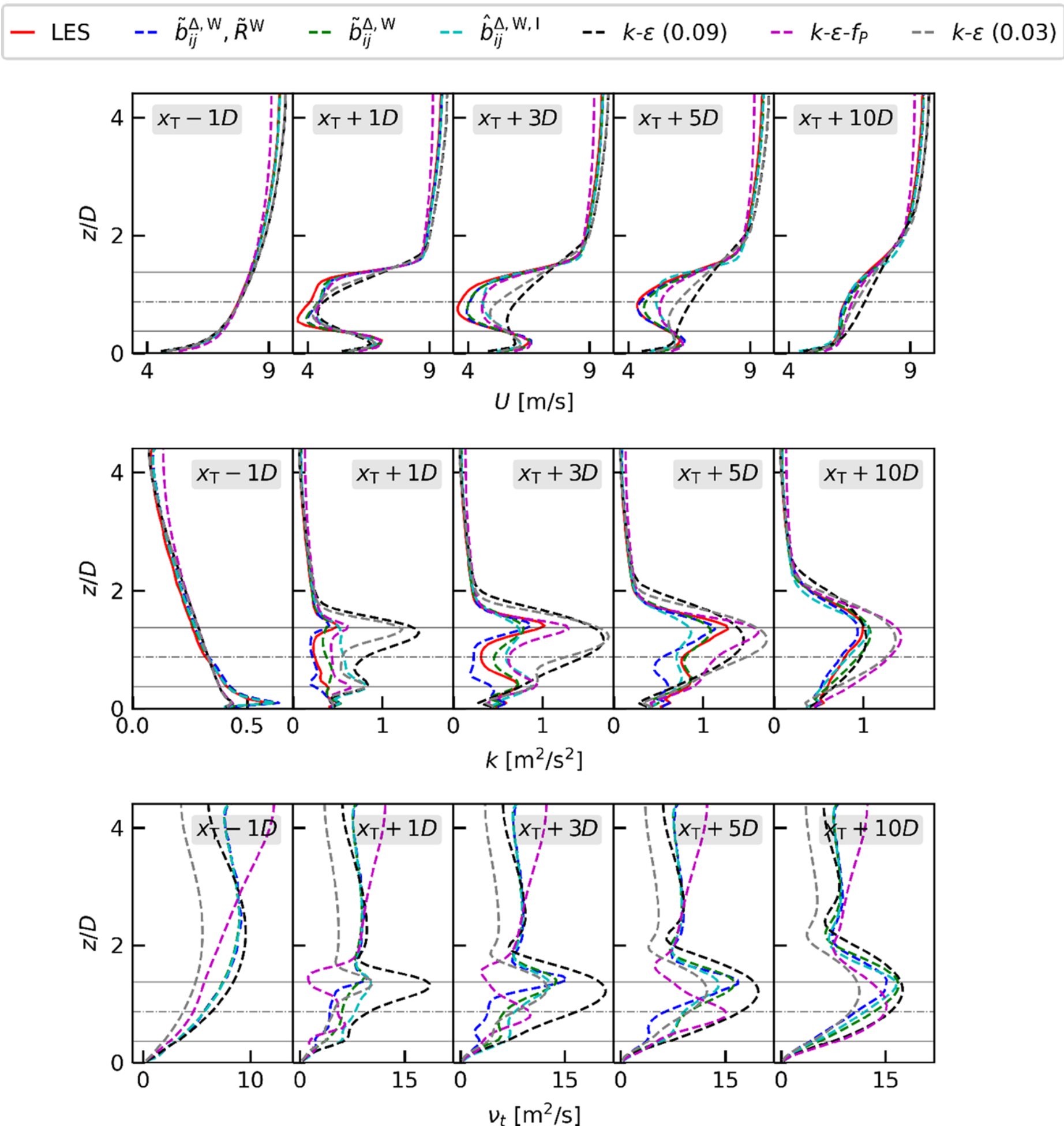

**Fig. 9** Comparison of LES and various RANS models for prediction of $U_x$, $k$ and $\nu_t$ profiles for the case 1T. The plots show the $x$-$z$ plane at $y = 2.5D$ for five different $x$ around the turbine position, $x_T$

#### 3.3.3 Summary

In summary, this section has demonstrated the impact of a modified eddy viscosity on flow prediction. The LES-derived optimal corrective terms showed substantial improvements over the baseline models, with the primary contribution coming from corrections to the deviatoric part of the RST. This finding supports the approach of focusing model development on $\tilde{b}_{ij}^{\Delta,\mathrm{W}}$. In addition, both the single-term model and the $k$-$\varepsilon$-$f_P$ model significantly improve $U_x$ predictions over the standard $k$-$\varepsilon$ model through eddy viscosity limiters that constrain turbulent mixing. Although neither model achieves $k$ predictions comparable to the LES data, their improved $U_x$ predictions demonstrate that eddy viscosity modifications to the standard $k$-$\varepsilon$ model can substantially enhance wake prediction performance in RANS simulations, even without accurate $k$ prediction.

### 3.4 Generalizability Assessment of the Data-Driven Model

In this section, we evaluate the single-term data-driven model's generalizability to explore its potential for prediction beyond the training case. The model is tested on the unseen six-turbine layouts with varying wake interactions (6T5D, 6T7D, and 6T7D-staggered, introduced in Sect. 2.3), despite being trained exclusively on a single-turbine case (1T).

Similar to the previous section, we first inspect the effectiveness of LES-derived corrective terms to verify whether the method extends to six-turbine cases. Following this, we evaluate the single-term model's performance on the test layouts by examining the $U_x$ and $k$ fields, as well as power production, primarily comparing with predictions from the $k$-$\varepsilon$ -$f_P$ model.

#### 3.4.1 Assessing the Effectiveness of the Corrective Terms

To evaluate the corrective terms, we examine the $U_x$ and $k$ fields for the 6T5D layout, which has the most intense wake interactions among the test cases. Figure 10 shows a top view at hub height, with the first column representing the $U_x$ field and the second column representing the $k$ field. Each row shows predictions from different models (including LES data), following the same order as Fig. 8.

The effectiveness of the corrective terms observed in the 1T case partially extends to the 6T5D layout. Similar to the 1T case, the full propagation and RST propagation simulations both produce $U_x$ fields that closely match the LES results, with substantially delayed wake recovery compared to the baseline $k$-$\varepsilon$ model ($C_\mu = 0.09$).

However, comparing the $k$ fields provides additional insight about the $\tilde{R}^{\mathrm{W}}$ term. Starting from the second turbine's wake, the RST propagation overpredicts $k$. This indicates that $\tilde{R}^{\mathrm{W}}$ is important for accurate $k$ prediction in multi-turbine configurations and highlights the need for a $k$ production correction, such as the one proposed in Zehtabiyan-Rezaie and Abkar (2024). Despite this, the RST correction alone still achieves significant improvement over the baseline, demonstrating the effectiveness of $\tilde{b}_{ij}^{\Delta,\mathrm{W}}$ even in multi-turbine layouts.

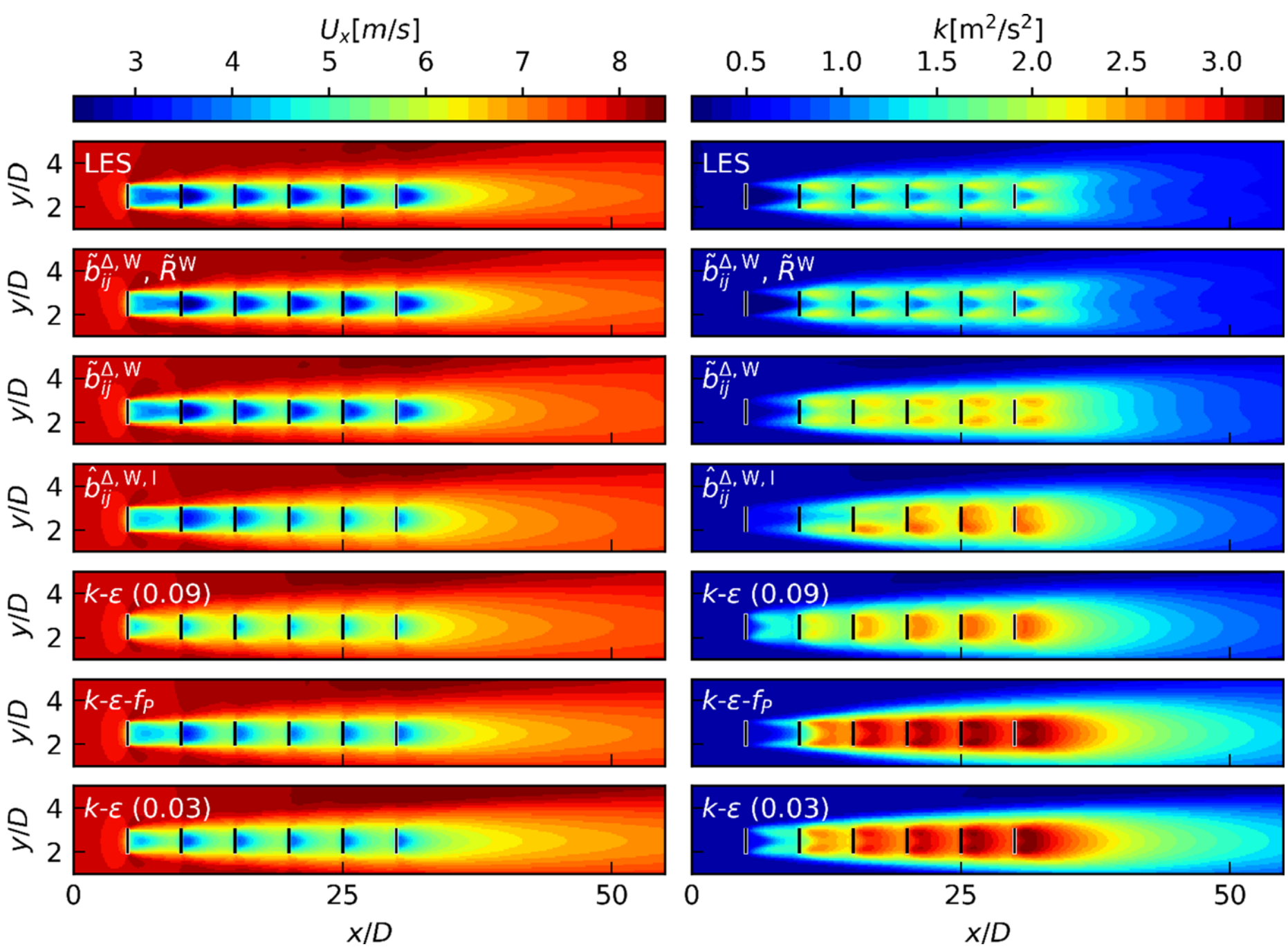

**Fig. 10** Comparison of LES and various RANS models for prediction of $U_x$ and $k$ fields for the case 6T5D. The plots show the x-y plane at the turbine hub height

Similar trends are also observed for the other layouts, with the 6T7D layout shown in Fig. 11 and the 6T7D-staggered layout shown in Fig. 12.

### 3.4.2 Assessing Data-Driven Model Generalizability on Test Layouts

To evaluate the generalizability of the single-term model, we compare its predictions with those of the $k$-$\varepsilon$-$f_P$ model across all test layouts (Figs. 10, 11 and 12). Throughout these simulations, both the single-term and $k$-$\varepsilon$-$f_P$ models remained stable, and no numerical issues were observed.

Both models exhibit improved wake recovery in the $U_x$ fields compared to their respective baselines, capturing the delayed recovery across all layouts. The most prominent differences are seen in the wakes of turbines experiencing the initial wake effects—specifically, the wake of the second turbine for the 6T5D and 6T7D cases, and the wake of the third turbine for the 6T7D-staggered case due to layout. This difference diminishes with increasing distance from the upstream turbine, with the 6T5D layout showing the most pronounced discrepancy and the 6T7D-staggered case the least. For detailed comparison, please see $U_x$ profiles sampled around the corresponding turbines for different layouts in plots in Figs. C1, C2 and C3 in App. C of the supplementary material. Beyond these minor variations, both models produce very similar $U_x$ fields.

With regard to the $k$ fields, notable discrepancies emerge between the models. The single-term model effectively reduces $k$ relative to its baseline, whereas the $k$-$\varepsilon$-$f_P$ model

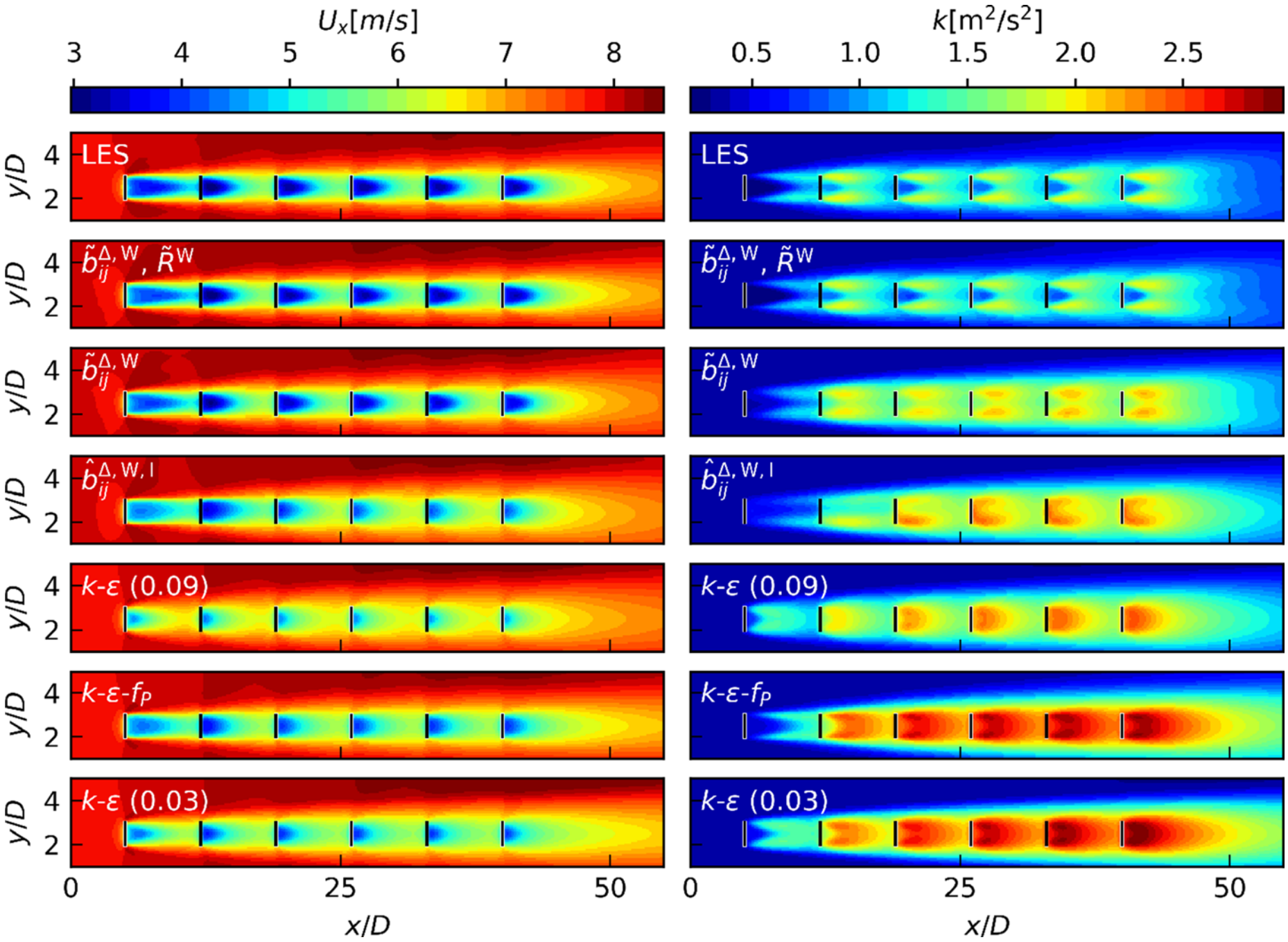

**Fig. 11** Comparison of LES and various RANS models for prediction of $U_x$ and $k$ fields for the case 6T7D. The plots show the $x$-$y$ plane at the turbine hub height

overestimates $k$. This observation aligns with findings from Sect. 3.2, where we attributed the discrepancy to differences in eddy viscosity limiter formulations. Nevertheless, although the models correct their corresponding baseline predictions to some extent, their $k$ prediction trends still largely follow those of the baseline models. This is clearly visible in the rotor-area-averaged normalized $k$ plot in Fig. C4 in App. C.

### 3.4.3 Power Prediction

To assess the data-driven model's practical utility for wind farm applications, we examine power output predictions across all layouts, as shown in Fig. 13.

The power predictions follow trends consistent with the $U_x$ fields, since turbine power is directly derived from $U_x$. Across all layouts, the single-term model exhibits moderate power overestimation for downstream turbines, while the $k$-$\varepsilon$-$f_P$ model achieves closer agreement with LES results.

A notable difference occurs in the 6T5D layout, where the single-term model underpredicts power for the third turbine. This stems from an overly delayed wake recovery behind the second turbine, as evident in Fig. 10 and Fig. C1. This underprediction only happens in the 6T5D layout, suggesting it occurs under stronger wake interactions—which are most pronounced in this closely-spaced arrangement.

Despite these variations, both models show substantial improvement over the baselines, which overpredict power across all turbines in all layouts. These results demonstrate that the data-driven model can generalize to predict wind farm power comparable to the established $k$-$\varepsilon$-$f_P$ model, and their similarity in velocity prediction across all cases is clearly visible in the rotor-area-averaged $U_x$ deficit plot in Fig. C5 in App. C.

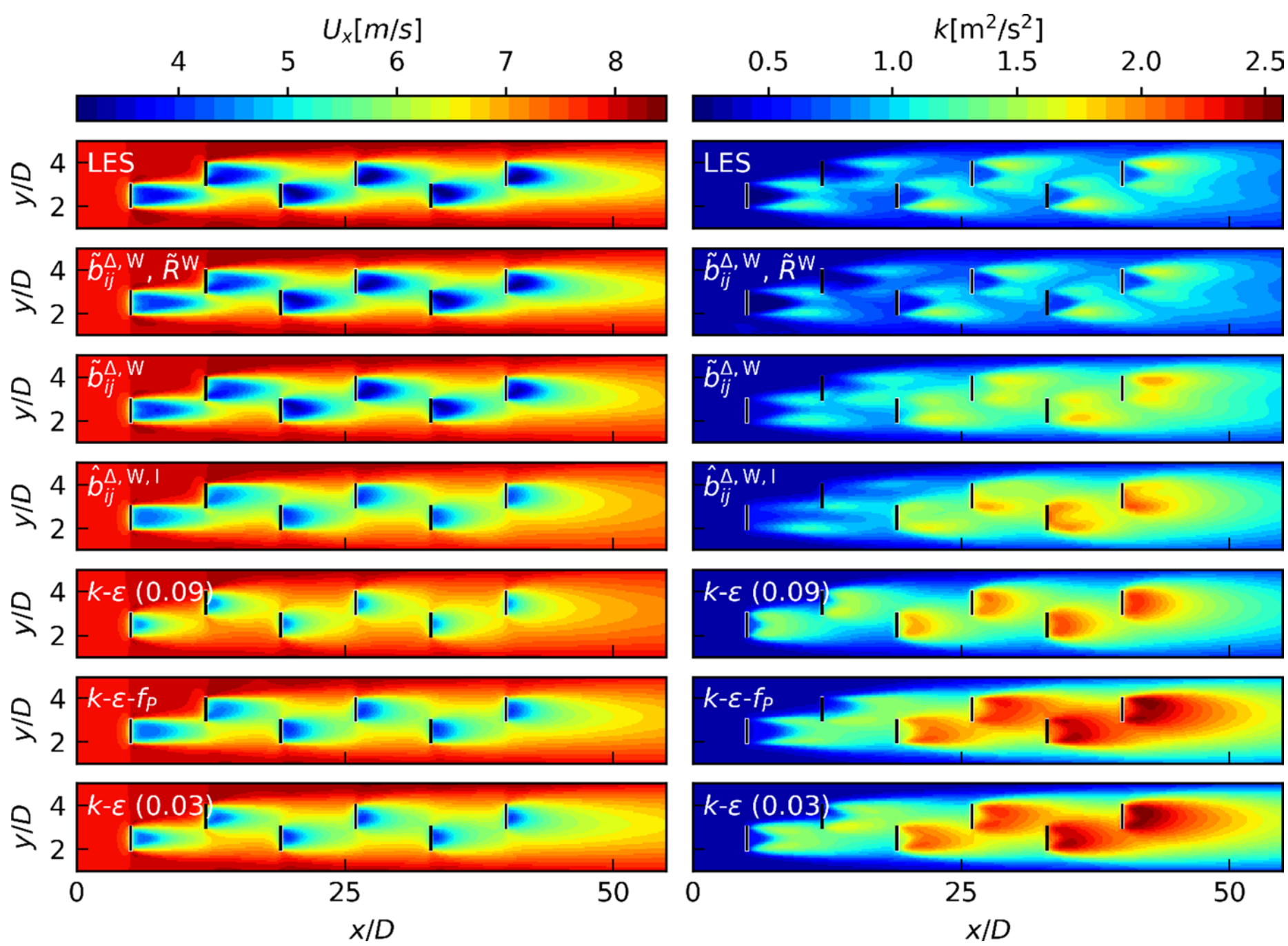

**Fig. 12** Comparison of LES and various RANS models for prediction of $U_x$ and $k$ fields for the case 6T7D-staggered. The plots show the $x$-$y$ plane at the turbine hub height

### 3.4.4 Summary

The results in this section demonstrate that the data-driven single-term model, which is trained only with single turbine cases, can predict comparable results for multi-turbine layouts to the established $k$-$\varepsilon$-$f_P$ model. Both models yield improved predictions of $U_x$ and power output compared to the baselines, with the $k$-$\varepsilon$-$f_P$ model achieving slightly better agreement with the LES. However, regarding the $k$ field, the single-term model shows better agreement with the LES, while the $k$-$\varepsilon$-$f_P$ model produces a notable overestimation following its baseline model behavior. Nonetheless, both models show significant improvements over their respective baselines, thereby demonstrating the generalization capability of not only the single-term model but also the $k$-$\varepsilon$-$f_P$ model in complex wind-farm layouts. Notably, no stability issues were encountered with either model during the simulations—implying that model simplicity may contribute to numerical stability.

Building on these findings, the similarity of the single-term model to the established $k$-$\varepsilon$-$f_P$ formulation, combined with its performance in the generalization study, suggests potential applicability under varying ABL conditions. The $k$-$\varepsilon$-$f_P$ model has demonstrated robust performance across a range of turbulence intensities (6–12.8%) and wind speeds

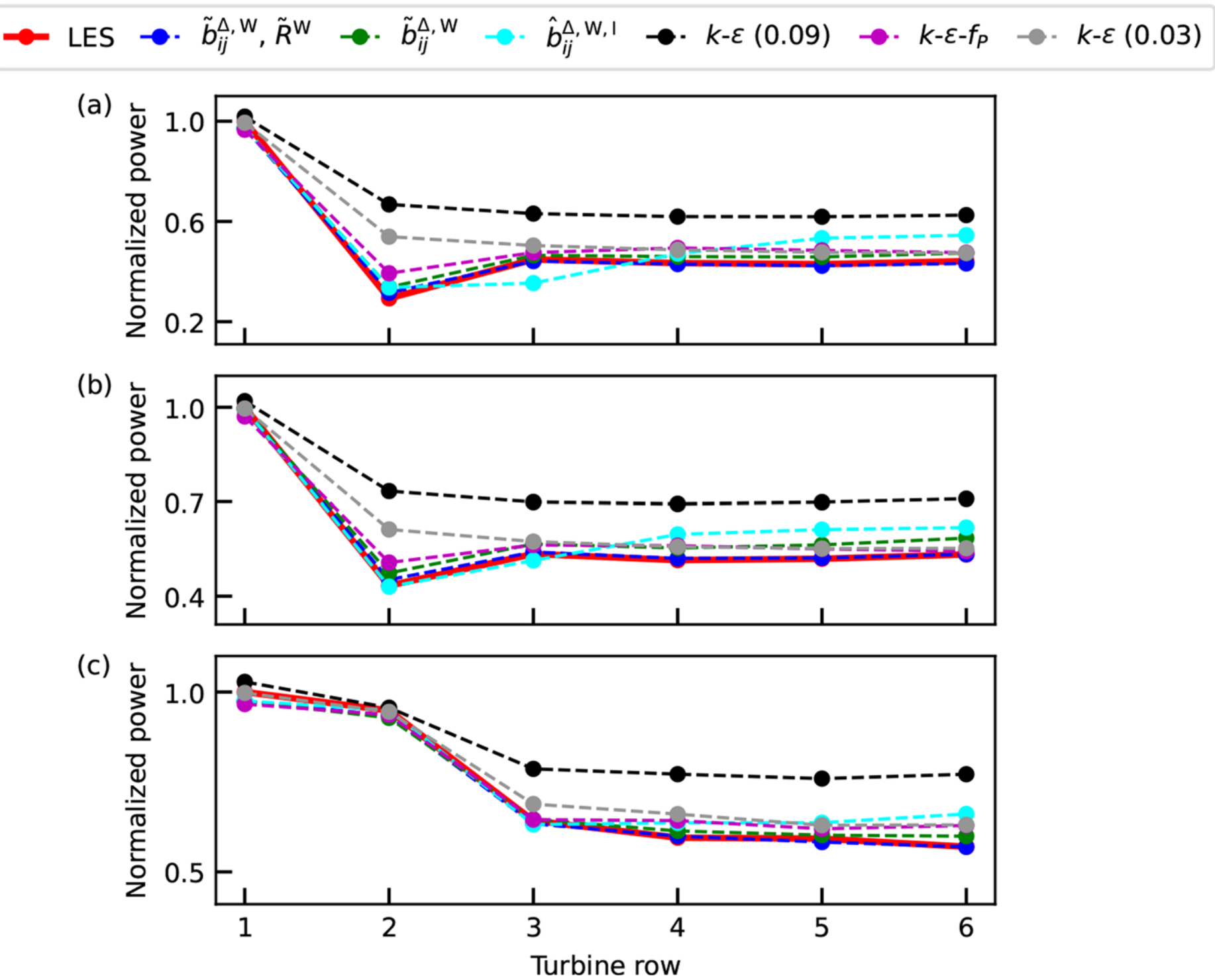

**Fig. 13** Comparison of LES and various RANS models for power prediction for the cases: (**a**) 6T5D, (**b**) 6T7D and (**c**) 6T7D-staggered. The values are normalized relative to the first turbine in the LES

(7.45–10.9 m/s) in full-scale wake simulations (van der Laan 2014), and the similar structure of the single-term model implies it could yield comparable results. Moreover, as shown in the generalization study, the model effectively predicted the wakes of downstream turbines subjected to modified inlet conditions resulting from upstream wakes. This supports its potential for broader application, although further validation with LES or experimental data is needed.

Furthermore, regarding the corrective term effectiveness, results highlight the role of $\tilde{R}^{\mathrm{W}}$ in obtaining improved $k$ fields for multi-turbine configurations. This supports the idea of having a $k$ production correction in the baseline model, as exemplified in Zehtabiyan-Rezaie and Abkar (2024).

## 4 Conclusion

This study presents a simple and interpretable data-driven turbulence model for wind turbine wake prediction, discovered through the SpaRTA framework using LES data from a single-turbine case.

The key finding of this work is that the symbolic regression method, namely SpaRTA, was able to discover an interpretable, explicit single-term model from high-fidelity data. The closed-form nature of the discovered model allows for direct analysis and analytical com-parison with existing models—capabilities that are typically not feasible with most black-box machine learning techniques. Moreover, the fact that a purely data-driven approach independently rediscovered a model structure closely resembling the $k$–$\varepsilon$–$f_P$ formulation is a compelling result, highlighting the potential of interpretable model discovery to reveal or validate underlying physical mechanisms directly from data.

Furthermore, the data-driven model demonstrates generalization capability across unseen six-turbine layouts with varying levels of wake interaction, despite being trained only on a single-turbine setup. In velocity field and power predictions, the model performs comparably to the $k$–$\varepsilon$–$f_P$ model, with the latter showing slightly better accuracy. However, the data-driven model shows better agreement with LES in predicting TKE, particularly by reducing the overestimation observed in the $k$–$\varepsilon$–$f_P$ model. Additionally, across all test cases, the data-driven model maintained numerical stability—a common challenge for complex data-driven approaches—suggesting that a simpler model may help promote numerical stability. These results, together with the model's structural similarity to the $k$–$\varepsilon$–$f_P$ formulation and its strong performance in generalization tests, suggest potential applicability under varying ABL conditions.

These findings demonstrate that the symbolic regression method can discover physically interpretable turbulence models that compete with traditional approaches.

Future work should address several areas for improvement. First, to overcome the limitation in TKE prediction, modeling of the $k$-equation corrective term is needed. This extension would fully exploit the SpaRTA framework. Second, systematic studies of the data-driven model's constants would help define its operational limits and applicability to various flow conditions. Finally, the current model cannot fully capture the anisotropy of the RST, as it belongs to the LEVM group. Hence, discovering a simple NLEVM with the SpaRTA framework remains a task for future work.

**Supplementary Information** The online version contains supplementary material available at https://doi.org/10.1007/s10494-025-00679-y.

**Acknowledgements** The authors express their appreciation to Julia Steiner for the RANS code and Mahdi Abkar for the LES dataset. The authors also thank the reviewers for their constructive feedback, which helped improve the quality of this manuscript. They are deeply grateful to the late Professor Dries Allaerts, whose invaluable support and guidance significantly contributed to this work.

**Author Contributions** K.J.: Conceptualization (equal); Methodology (equal); Coding (lead); Formal analysis (lead); Writing – original draft (lead). A.E.: Methodology (supporting); Writing – review and editing (equal). N.A.K.D.: Supervision (equal), Methodology (supporting); Writing – review and editing (equal). R.P.D.: Conceptualization (equal); Methodology (equal); Supervision (equal); Writing – review and editing (equal).

**Funding** A. E. acknowledges the support from the GO-VIKING project (Grant No. 101,060,826), funded by the Euratom Research and Training Programme of the European Union. N. A. K. D. acknowledges support by ERC grant (CONTEXT, GA no. 101,161,294).

**Data Availability** The data used in the current study are available from the corresponding author on reasonable request.

## Declarations

**Consent to Publish** Not applicable.

**Competing Interests** The authors declare no competing interests.